# AN EXPERIMENTAL EVALUATION OF ANALOG TV CABLE SERVICES DISTRIBUTED IN GPON ARCHITECTURE

Radu ARSINTE , Eugen LUPU
*Technical University of Cluj-Napoca, Communications Department, Romania*
*Str. Baritiu 26-28,Radu.Arsinte{Eugen.Lupu}@com.utcluj.ro*

**Abstract**: This paper explores the benefits of GPON (Gigabit Passive Optical Networks) in analog TV services. Analog TV service is still present in the standard triple play distribution architectures, as an effect of unique advantages: simple distribution in an apartment via standard RF splitters, unlimited number of viewing sites, real time behavior by lack of encoding/decoding processes. Of course, the quality is still limited by analogic standards, but the price/performance ratio is unbeatable. The most important parameters characterizing the analog TV performance are described, the network architecture is emphasized. To have a reference, the test architecture is a sub-network of a commercial telecom company. In this case only the quality of TV reception is performed. Using open-source software, and the integrated TV tuner board makes possible to test the levels and S/N ratio for all the analog TV channels of the spectrum. The system opens the possibility to do the same for digital channels in DVB-C standard, with minimal changes.

## I. INTRODUCTION

### A. History [1]

The PON technology was launched for the first time in the 1990s of the last century, when several major European operators, including British Telecom and France Telecom, joined together in a consortium to develop the technology of multiple access over a single fiber. This is how the technology is, the distinguishing feature of which is to combine the use of passive optical splitters traffic (splitters), do not require powering and complex servicing. In the first stage, the GPON cost has limited the number of installation points to business sites (offices, industrial buildings).

Figure 1 present three possible architectures in residential access. Today GPON (PON) (c) allows a fiber-optic cable to the apartment and provide a bandwidth of up to 1 Gbit/s, which is 100 times faster than ADSL-access (a or b). Implementing technologies based on GPON (PON) ensure the provision of Triple-Play services over a single fiber: Internet, VoIP, IPTV; and high speed transmission providing the ability to carry and view multiple HDTV channels simultaneously. The new network allows you to make phone calls using IP-telephony technology.

### B. Characteristics [1]

The technology offers, in the basic version, High-speed Internet access at speeds up to 1 GB/s. The latest interactive IP-TV, which might give an opportunity not only to watch more than 100 satellite and terrestrial digital TV channels, including at least 30% channels of high-resolution format HDTV, but use also interactive services: order record movie and view it after broadcasting the show, view telecast, broadcast of live programs, order a movie from the video library quality communications (VOD) with the ability to connect both conventional and IP telephony with advanced features, unlimited numbers on one line, and save the number when moving. Using IP-telephony services can significantly save on calls to other cities.

Unlike other operators providing the optical path technology to the building only, GPON (PON) implies fiber channel directly to the subscriber's apartment. This not only significantly improve the quality of signal transmission (data video, voice), but increase by dozens of times the data rate available over the network to transmit several TV channels simultaneously in High Definition.

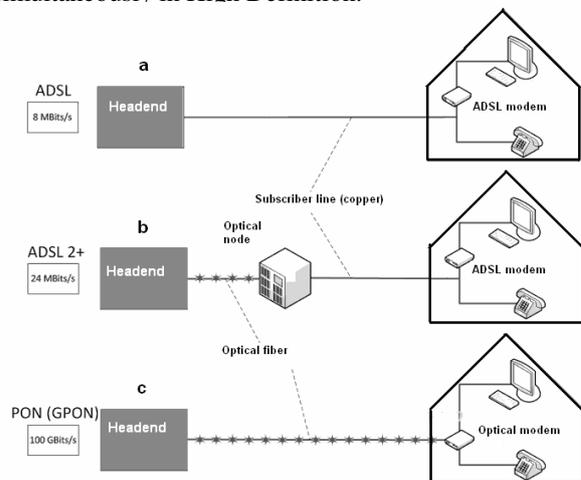

*Figure.1. A comparison of complex fixed residential data access technologies*

### C. Advantages of GPON (PON)

There are few keywords and concepts describing the most important advantages.

*Reliability* – it is protected from unauthorized access, fiber-optic cable is insensitive to electromagnetic influences;

*Speed* - the optical fiber has a huge bandwidth, so the speed and quality of transmission is superior over other technologies (both wired and wireless);

*Additional services*, creating the possibility in the future, on the basis of the GPON technology, to connect the future

___

services, remote surveillance of any site (house, office) and the possibility of the smooth operation; without affecting the performance of existing services.
*Complexity* - PON technology allows you to connect to the fiber at least three services - Internet, digital TV, telephone;
*Flexibility* - GPON (PON) technology allows adjustment of the equipment in accordance with the individual characteristics of the subscriber and provides exactly the level of service required by the subscriber;
*Improvisational* - it allows you to use the most modern technologies of data transmission;
*Environmentally friendly* - it does not have electrical/magnetic radiation;
*Provision for future expansion* - at constant growth requirements of content transmitted over the network speeding the transfer solves this problem for several years ahead.

### D. GPON General Architecture [2]

OLT (Figure2) is placed in the central office or in a technical room of the residential area. The splitter is placed outside a building, usually in a central residential area, where it then goes up to every fiber distribution housing, terminating at ONU/ONT (Optical Network Unit/Terminal) situated therein. In fact ONU/OUT are similar devices: ONT is an ITU-T term, whereas ONU is an IEEE term. The services provided by such a network include VoIP, IPTV, Monitor, HSIA (High Speed Internet access), CATV, etc. ONU interfaces are of the type: POTS, FE / GE, WIFI, RF, etc. The bandwidth provided to each subscriber is relative, dependent on the number of ONUs, in the 10-1000Mbps range.

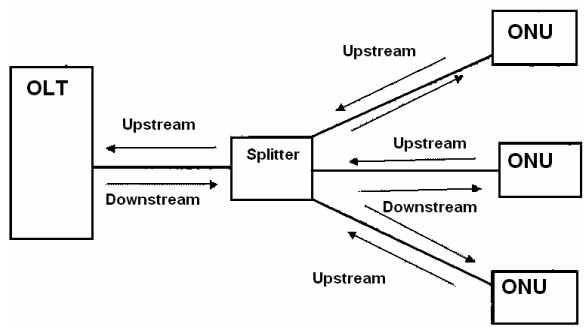

*Figure 2. FTTH topology*

## II. TEST ARCHITECTURE

We are using as a test architecture a sub-network of a commercial telecom company. The company offers a full set of services: analog and digital cable TV, VoIP telephony and Internet access (figure 3). The future expansion is provided as an interactive IPTV service, but it is not included in the present test sequence.

It is well known that traditional cable TV operators ([3], [4]) are offering all the services in a common architecture, but GPON brings a novel approach, with at least the same advantages.

In this approach the subscriber is provided with all the resources of an optical cable that is placed directly into his apartment, in contrast to the operators of other home networks where the channel allocated to the house and, accordingly, is equally divided between the connected users. The network tested is based on a Huawei HG8247H GPON Terminal [5], allowing combined Internet, telephony and cable TV services. Additionally, the terminal delivers WiFi services (Access Point).

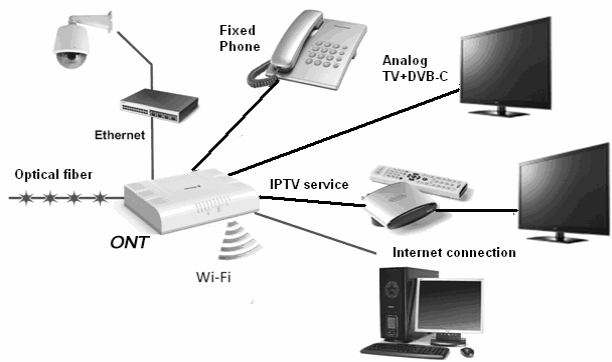

*Figure 3.In house architecture of GPON based services*

### A. Procedure and Software

The methods to test analog TV [6] are already a classical matter. The drawback of it is, that it is quite time consuming and expensive to conduct. To solve this problem, objective measurements are used; measurements performed using a computer or networked instruments. The most well known and most widely used in the context of video coding is the so-called Peak-SNR or PSNR. Computing the mean squared error between the original and the decoded video or image as MSE (average of all squared differences for the pixels of original and decoded version), then the PSNR is computed as:

$$PSNR = 10 \cdot \log_{10}\left[\frac{(2^n - 1)^2}{MSE}\right] \quad (1)$$

Where n is the number of bits per pixel value (usually 8-10). There $2^n-1$ represents the peak value in the image (usually 255), and the name "Peak" SNR (the calculation is done per channel, i.e. for R, G, B or Y,Cb,Cr).

The normal procedure requires reference images to be broadcasted on the desired channel, difficult to implement in a real commercial network, where the normal broadcast can not be interrupted for tests.

The employed software uses the TV tuner board to test and measure the S/N ratio of analog TV channels in a defined frequency domain.

The measurement is performed on the Vertical Blanking Interval (VBI) the only part of TV signal with a defined level, the normal (active) part of TV signal having a random structure, depending on the displayed signal.

In the measurement of the SNR it is assumed that the input signal does not change, and is constant on the SNR measurement period. This makes it easy to separate the useful signal from noise. To determine the value of the desired signal uses the fact that the expectation (arithmetic mean values of the signal samples, electronics expectation corresponds to a constant component of the signal) of the noise is equal to 0. In this case, the arithmetic mean of the resulting amount of noise signals and the desired signal (ie. E). It what we get from the output of the video decoder) will correspond to the value of the original useful signal, that is, to what we have agreed, that the useful signal is a constant component. It should be noted that to calculate the value accurately of the original desired signal requires a large number of samples - at least several dozen. As mentioned above, the expectation of the noise and, hence, its DC component are equal to zero. It allows you to calculate the actual noise value as the value of the standard deviation. In other words, the current noise signal value is expressed as

$$V_N = \sqrt{\frac{1}{N-1}\sum_{i=0}^{N-1}(x_i - V_{REF})^2} \quad (2)$$

where $V_N$ - rms of the noise, $V_{REF}$ - value of the signal on the VBI, $x_i$ – the i-th value of the digitized reference signal of the SNR measurement period, i going from 0 to N-1.

Standard formula used to determine the SNR:

$$SNR(dB) = 20 \bullet \log\left(\frac{A_{FS}}{V_N}\right) \quad (3)$$

where $V_{FS}$ – is the maximum amplitude of the signal and is considered to be the difference between 16 and 235 (in this range of the original signal values are delivered the values of the digitized signal, according to ITU-R 601), i.e. 219, $V_N$ - active noise value calculated in the previous step.

It is obvious that all the calculations are using the statistical properties of the signal. However, as is known, these methods give certain error. Therefore, the captured instantaneous SNR corresponds to the real value of the SNR with some accuracy. The margin of error in this case is given by:

$$E = \frac{V_N}{\sqrt{N}} \quad (4)$$

where $V_N$ - RMS value of the noise, N - number of samples, E - statistical error.

### B. Hardware

As described in [7] and [8] an integrated TV tuner board could be used as a test instrument, at least after a calibration procedure as presented there. In any case, the system can be used to establish the stability of the TV cable system over the entire frequency domain. This is normally sufficient for field engineers or the skilful users.

The site [7] contains the drives and implementation for the two main chipsets used in PCI based TV tuners – BT878 (Conexant) and SAA713x (Philips). This is one of the limitations of the proposed methodology, and the adaptation to the new generations of chipsets is not tested yet. The precision of ADC is at least 9 bit, sufficient for quick evaluation purposes.

### III. EXPERIMENTAL RESULTS
#### A. Introduction

In principle the software should work with any arbitrary "TV tuner card", delivering VBI information through Direct Show at eventually different sample frequency of the ADC, resulting in a different number of samples per line, but with no additional information, basically being necessary to deliver the area to calculate SNR in graphical viewer on an empty black line (so there is no teletext, measuring lines or the program identifiers, etc.) through the window text interface.

The quality of the TV tuner board, do influence the precision of the measurement, and a procedure of self evaluation as described in [8] could be necessary. In general, modern TV tuner boards have sufficient SNR provision (more than 48-50dB [9]) to be used as reference, in cable networks with a SNR below 42-45dB. The program allows modifying the parameters of the measurement procedure in order to accommodate different configurations and experimental setup. To compensate the relatively low number of measurements on each frame (on VBI only), a series of measurements (up to 30 frames) can be realized with the accumulation of multiple frames.

### B. Experimental setup

The experimental setup is realized starting from the GPON terminal TV cable output and distributed via a RF splitter as presented in figure 4.

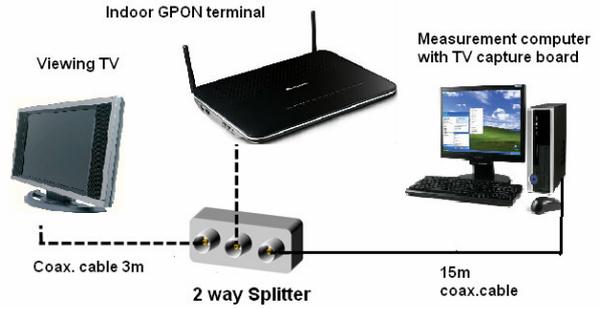

*Figure 4. Experimental setup for SNR measurement*

*Table 1.*

| No. | Channel name | Ch. | Freq. | No. | Channel name | Ch. | Freq |
|---|---|---|---|---|---|---|---|
| 1 | TVR1 | S02 | 112.25 | 9 | ProTV | C06 | 182.25 |
| 2 | TVR2 | S03 | 119.25 | 10 | Antena1 | C07 | 189.25 |
| 3 | TVR3 | S04 | 126.25 | 11 | KanalD | C08 | 196.25 |
| 4 | TV5Monde | S05 | 133.25 | 12 | Prima | C09 | 203.25 |
| 5 | TVR Cluj | S07 | 147.25 | 13 | Nat. TV | C12 | 224.25 |
| 6 | Antena1 loc | S08 | 154.25 | 14 | Stars | S11 | 231.25 |
| 7 | Un. Carrier | S09 | 161.25 | 15 | Gold | S12 | 238.25 |
| 8 | ETV | S10 | 168.25 | 16 | Acasa | S13 | 245.25 |

The number of channels in the investigated analog TV cable network is over 55, and only a partial list is presented in Table 1.

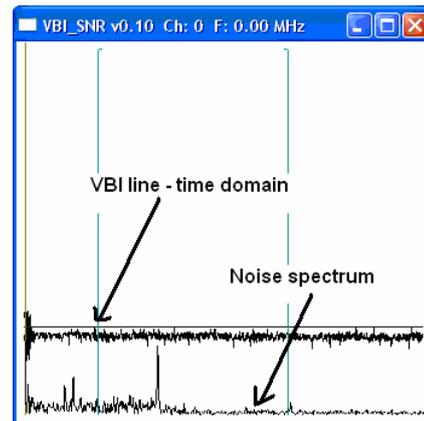

*Figure 5. Graphical interface of VBI_SNR analyzer[7]*

The tools able to measure the signal are very complex, involving the black level extraction from "clean" VBI lines and SNR evaluation using the described procedure (previous paragraph). A full set of measurement tools, presented simultaneously on the screen is composed of the following program utilities:

- VBI_SNR_BT8x8
- PlayTune

**VBI_SNR** is the measurement program, different for Bt878 or SAA713x TV tuner versions. In our case we used the 878 (Conexant) version.

The graphical user interface is presented in figure 5. The upper curve represents the evolution in time domain of the selected VBI line (amplitude), the lower curve representing the frequency spectrum of the signal corresponding to the same line. This unusual representation makes possible to observe simultaneously the noise in time domain and in the frequency spectrum. Attached to the graphical view is a text interface (figure 6), allowing to change the main parameters of the analysis.

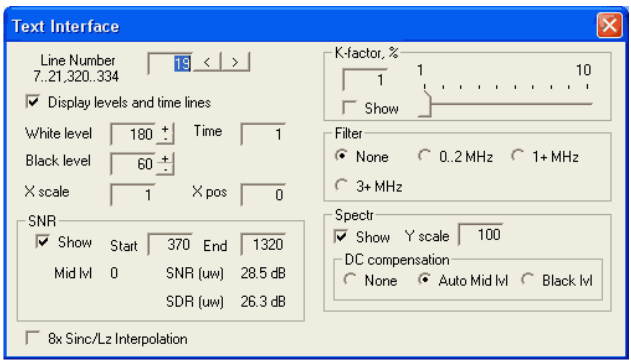

*Figure 6. Text interface of the SNR analyzer [7]*

**PlayTune** displays the video content of selected channels, to ensure the user that the measurements are performed on a stable (synchronized) channel.

**C. Results**

Table 2 presents the results of SNR measurement for few channels in VHF domain.

*Table 2.*

| Channel | S02 | S03 | S04 | S05 | S07 | S08 |
|---|---|---|---|---|---|---|
| SNR1[dB] | 29.4 | 30.6 | 30.5 | 36.4 | 33.3 | 29.6 |
| SNR2 [dB] | 40.1 | 39.7 | 40.4 | 39.8 | 40.7 | 39.7 |

| Channel | S09 | S10 | C06 | C07 | C08 | C09 |
|---|---|---|---|---|---|---|
| SNR1[dB] | 28.8 | 36.6 | 31.4 | 34.3 | 32.2 | 36.4 |
| SNR2 [dB] | 39.8 | 40.1 | 40.7 | 40.2 | 39.9 | 40.3 |

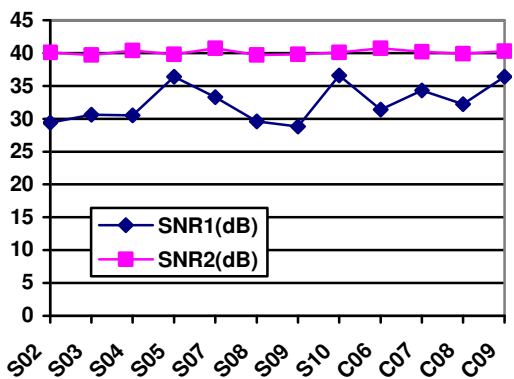

*Figure 7. Variation of Unfiltered (SNR1) and filtered (SNR2- LPF 2Mhz) SNR in S02…C09 (112…203MHz) domain [in dB]*

The SNR measurements can be realized using the full spectrum of the noise, or only filtered components (see figure 6 - option filter) can be used in calculations. The importance of this option is presented in figure 7 where unfiltered and filtered measurements are presented for the same programs from Table 2 of the analog TV spectrum. Some channels are not occupied, as a precaution from the cable operator, to avoid aerial frequencies used in the same geographical area.

It is easy to observe that the unfiltered measurement is affected by the RF disturbances in the area, resulting in a variable SNR in the frequency domain, while the filtered measurements are constantly in the 40dB SNR domain. Fortunately, the detected "noise" includes also additional carriers, like the sound carrier, and the TV set filters the additional carriers before processing the video. An additional possible explanation is the relative long distribution coax cable (15m) susceptible to collect disturbances form the noisy urban RF environment.

### IV. CONCLUSIONS

In this paper is described a test architecture able to evaluate the behavior and the quality of Analog TV distribution in GPON networks. The system is based on a PC (in desktop or laptop architecture) with integrated TV tuner (analog or hybrid), and an open source software package, able to scan and measure the level and S/R ratio of each channel. The precision of the evaluation is sufficient for field and service purposes, but the methodology is full applicable to professional standards with a better and calibrated acquisition board.

In this moment the evaluation is extended in UHF domain to evaluate the efficiency of the evaluation in the higher area of the TV spectrum. An extension to Digital TV (DVB-C) evaluation is previewed.

The system is dedicated mainly for educational purposes, the open source choice allowing changes and customization of the evaluation procedure. A major drawback is the (still) presence of some software bugs, but once detected, they will be eliminated.